# Flexibility-constrained Operation Scheduling of Active Distribution Networks with Microgrids

Sarah Allahmoradi, *Student Member, IEEE,* Mohsen Parsa Moghaddam, *Senior Member, IEEE*, Salah Bahramara, *Member, IEEE*, and Pouria Sheikhahmadi

*Abstract*— Regarding the variability of renewable energy sources (RESs), especially in the operation time periods, their high penetration faces the net load pattern of the power system with a major ramp rate challenge. Although employing a market-based mechanism by the independent system operator (ISO) can address this challenge, it may not be possible to handle this challenge in all networks. In such networks, the required flexibility can be supplied through decreasing the ramp rate of the purchased power of distribution companies (Discos) from the market since their net load has an important impact on the system's net load. Therefore, a flexibility-constrained operation problem for the distribution networks with distributed energy resources (DERs) and microgrids (MGs) is proposed in this paper to decrease the ramp-rate of the Disco's purchased power from the market. This problem is formulated using a bi-level two-stage stochastic model where the problem of the Disco and the MGs are modeled as the upper-level and lower level problems, respectively. The proposed model is applied to the IEEE 33-bus standard test network with three MGs. The results show the effectiveness of the proposed model to decrease the ramp-rate of the Disco's purchased power from the market.

## NOMENCLATURE

*Indices and sets:*

| | |
|---|---|
| $i, h$ | Index for buses |
| $j/J$ | Index/Set of MGs |
| $g/G$ | Index/Set of DGs |
| $s/S$ | Index/Set of scenarios |
| $t/T$ | Index/Set of time |
| $k/K$ | Index/Set of PVs |
| $L$ | Set of grid lines |
| $M$ | Set of busses that have MG |
| $O$ | Set of busses that have DG |
| $P$ | Set of busses that have PV |
| $N$ | Sets of buses |

*Parameters:*

| | |
|---|---|
| $R_{i,h}/Z_{i,h}$ | Line resistance/ line impedance (Ω) |
| $\overline{V}_i/\underline{V}_i$ | Maximum/ minimum voltage magnitude (V) |
| $\overline{I}_{j,h}$ | Maximum current flow magnitude (A) |
| $P_{i,t,s}^D$ | Distribution network load (DNL) (MW) |
| $\overline{P}^{IL,D}$ | Maximum load curtailment of DNL (MW) |
| $C_j^{DG}$ | Bid of DG operated by MG ($/MWh) |
| $C_{j,t}^{IL}$ | Bid of IL operated by MG ($/MWh) |
| $E_j^{ini}$ | Initial amount of energy storage (ES) (MWh) |
| $\overline{E}_j/\underline{E}_j$ | Maximum/minimum amount of ES (MWh) |
| $\overline{P}_j^{ES}$ | Maximum charging/discharging power of ES (MW) |
| $P_{j,t}^{demand}$ | MGs' demand (MW) |
| $P_j^{DG,ini}$ | Initial power generation of DG (MW) |
| $\overline{P}_j^{DG}/\underline{P}_j^{DG}$ | Maximum/minimum power generation of DG (MW) |
| $\overline{P}^E$ | Maximum purchased power from wholesale energy market (WEM) (MW) |
| $\overline{P}_j^{IL}$ | Maximum MG's load curtailment (MW) |
| $\overline{P}_j^{MG}$ | Maximum of power exchange between Disco and MGs (MW) |
| $P_{k,i,t,s}^{PV}/P_{j,t}^{PV}$ | Power generation of PV operated by Disco/MGs (MW) |
| $RDN_j^{DG}/RUP_j^{DG}$ | Ramp down/up rate of DGs (MW/h) |
| $\rho_t^{WEM}$ | WEM price ($/MWh) |
| $\rho_t^D$ | Selling price to DNL ($/MWh) |
| $\rho_g^{DG,D}$ | Bid of DG operated by Disco ($/MWh) |
| $\rho_t^{IL,D}$ | Bid of IL operated by Disco ($/MWh) |
| $\rho_t^F$ | Penalty offer by the ISO ($/MWh) |
| $\overline{\rho}^{LEM}$ | Maximum price for power exchange between Disco and MGs ($/MWh) |
| $\eta_j^{ch}/\eta_j^{dch}$ | Charging/ discharging efficiency of ES (%) |
| $\pi_s$ | Probability of scenarios (%) |

*Variables:*

| | |
|---|---|
| $P_{i,h,t,s}^{fm}$ | Power flow leaves node i towards node h (MW) |
| $P_{i,h,t,s}^{to}$ | Power flow leaves node h toward node i (MW) |
| $P_{i,h,t,s}^{loss}$ | Power loss between nodes i and h (MW) |
| $V_{i,t,s}/V_{i,t,s}^2$ | Voltage magnitude and its square (V) |
| $I_{i,h,t,s}/I_{i,h,t,s}^2$ | Current flow magnitude and its square (A) |
| $E_{j,t}$ | The energy stored in ES (MWh) |
| $P_{i,t}^E$ | Purchased power by Disco from WEM (MW) |
| $P_{g,t,s}^{DG,D}$ | DG's power operated by Disco (MW) |
| $P_{i,t,s}^{IL,D}$ | The amount of IL operated by Disco (MW) |

| $P_{j,i,t}^{MG}$ | MGs' purchase/sell power from/to Disco (MW) |
| --- | --- |
| $P_{j,t}^{ch}/P_{j,t}^{dch}$ | Charging/discharging power of ES (MW) |
| $P_{j,t}^{DG}$ | DG's power operated by MGs (MW) |
| $P_{j,t}^{IL}$ | The amount of IL operated by MGs (MW) |
| $\rho_t^{LEM}$ | Price for power exchange between Disco and MGs ($/MWh) |
| $\Delta_t^F/\Delta_{j,t}^{MG}$ | Ramp of Disco/ ramp of MGs (MW/h) |

## I. INTRODUCTION

The penetration of renewable energy sources (RESs), such as wind power and photovoltaic (PV) systems, in the electrical energy systems, is increasing due to the environmental concerns and national goals for decarbonizing such systems [1]. For instance, several countries, such as Germany, Belgium, Sweden, Denmark, and the United States (Hawaii and California), have set ambitious targets to operate 100% renewable energy by 2050 [2]. The RESs bring some challenges to the energy systems regarding their uncertainties (unpredicted generation) [3] and variability (generation continually changes). The power generation pattern of the RESs, especially the PV systems changes the net load profile of the network. To be more specific, the 4 highest amount of the PV system generation around mid-day and decreasing their generation during the early evening besides increasing the residential load at this time drop the daily net load at noon and increase it significantly in the evening. As reported by the California Independent System Operator (CAISO), it needs 9235 MW ramp up in three hours between 16:10 and 19:10 on 24th July 2020 [4]. Independent system operators (ISOs) can tackle the intermittent behavior of the net load through the load following strategy with fast ramping units in timescales of minutes to hours. The penetration of RESs increases the need for ramping and load-following services in the power systems [5]. Although the traditional approaches can be used to address this ramping, they might lead to an increase in the operation cost and the greenhouse gas emissions, or they may require new investments or even become limited by the possible network congestion. Therefore, the ISO needs new approaches to meet the required ramp of the system to mitigate the problems related to the high penetration of RESs in the power systems.

Flexible ramping products (FRPs) are introduced as a solution for this problem where a market mechanism is developed by the ISO to optimally dispatch the flexible energy resources. In this mechanism, regarding the required amount of the system's ramp and the offers of the energy market players, the ISO clears the flexibility market to meet its required ramp in the real-time (RT) operation. This problem is addressed in the literature from two different points of view; 1) modeling the FRPs market besides the energy and the reserve markets which are cleared by the ISO [6, 7] and 2) modeling the decision-making problem of the energy players to provide flexibility to the system besides participating in other markets [8-10]. The co-optimization of the energy and the FRPs is modeled in [6] using the Nash Cournot approach. A multi-stage stochastic bi-level model is developed in [7] to model the competition of the generation companies (Gencos) in the day-ahead (DA), RT, and FRPs 5 markets. The decision-making problem of a MG in the wholesale energy, reserve, and FRPs is modeled using a hybrid stochastic/robust optimization to address the uncertainties of the output power of RESs and the energy prices in [8]. The operation problem of an energy hub to meet its required energy is modeled when it participates in the DA, RT, and FRPs markets as a price-taker player [9]. Providing the FRPs to the system besides participating in the DA and reserve markets by a battery energy storage aggregator is addressed in [10].

Either not being able to implement the market-based mechanism in some systems [11], or lacking enough energy resources to provide the flexibility requirement of the system or even both of which are the reasons that enable the ISO to decrease the amount of the flexibility requirement in such systems. This reduction can be done by decreasing the ramp rate of the system's net load, especially in the distribution network (DN) level using the capability of the distributed energy resources (DERs). The reason for this management is that the net load of the DNs has an important impact on the net load of the system since the high amount of RESs are connected to this network and the distribution company (Disco) purchases a high amount of power from the wholesale energy markets (WEMs). The issue of decreasing the net load's ramp rate to increase the flexibility of the system is addressed in [11-13]. In these studies, a power exchange-based mechanism is proposed where both inter-hour and intra-hour net load variability of the DN decrease with adding the flexibility constraints in the MG operation problem. In this framework, the ramp rate limitations of the net load are defined by the grid operator to obtain the desired net load of the DN [13]. Since the MGs are equipped with different DERs, and they are located in different locations of the network, defining a fixed amount for the ramp limitations for the MGs cannot be perfectly helpful in decreasing the ramp of the DNs with a high number of MGs. Alternatively, in this paper, the amount of reduction the ramp rate of the DN's net load is modeled as a decision variable of the Disco where it is determined through the optimal decisions of the Disco to trade power with the 6 MGs and purchasing power from the distributed generations (DGs) and Interruptible Loads (ILs) in its network. To encourage the Discos to decrease their net load's ramp rate in such framework, a compensatory payment mechanism is defined in this paper which is described in the next section. The proposed decision-making framework of the Disco is formulated as a flexible-constrained two-stage stochastic bi-level optimization approach. In this model, the operation problem of the Disco and the MGs are modeled as the upper-level (UL) and the lower-level (LL) problems. Moreover, to address the uncertainties of the PV system and the demand of the DN, the operation problem of the Disco is formulated as a two-stage stochastic optimization approach. Therefore, the main contributions of this paper are as follows:

- Proposing a flexibility-constrained operation problem for the DNs in the presence of DERs and the MGs in order to decrease the ramp of the purchased power of the Disco from the WEM,
- Proposing a bi-level two-stage stochastic model to formulate the flexibility constrained decision-making of the Disco and the MGs considering the uncertainties of

PV system and demand.

The rest of this paper is organized as follows. Section II presents the problem description. In section III, the proposed model is formulated. The numerical results are analyzed in section IV. Finally, section VI concludes the paper.

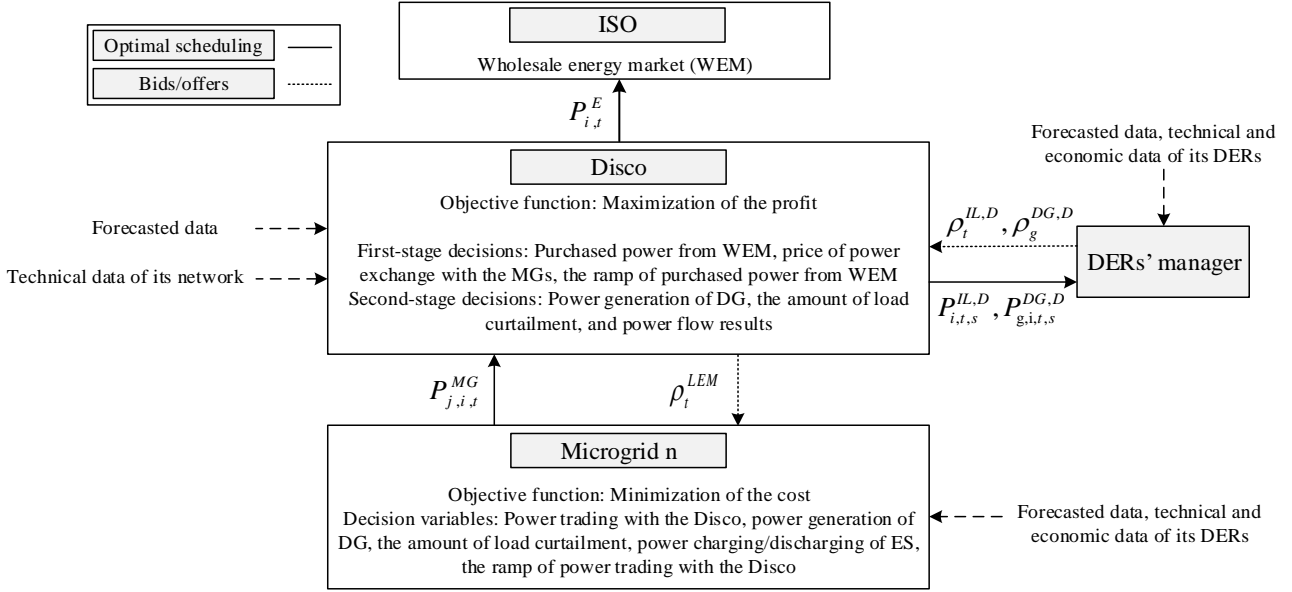

**Fig. 1.** Decision-making framework to participate in the LEM and the WEM

## II. PROBLEM DESCRIPTION

The proposed decision-making framework of a Disco to provide flexibility to the WEM when it trades energy with the MGs and the DERs owners in its network is described in Fig. 1. Providing flexibility is done through decreasing the ramp rate of the Disco's purchased power from the WEM. To model this framework, a bi-level optimization approach is used where the operation problem of the Disco and the MGs are modeled as the upper-level (UL) and lower-level (LL) problems, respectively. The aim of the Disco is to maximize its profit through the purchased power from the WEM, the scheduling of the DGs and ILs in its networks, and the trading power with the MGs. These decision variables of the Disco are determined using its required input data consisting of the bids and the technical constraints from the DGs and IL, the penalty bids sent from the ISO, and the forecast data and the technical data of its network. Regarding the uncertainties of the output power of PV systems and the demand, the decision-making problem of the Disco is modeled as a two-stage stochastic programming. The first-stage (here and now) decisions are independent of scenario decisions, while the second stage (wait and see) decisions are made after the occurrence of each scenario. In the proposed model, the first-stage variables are the purchased power from the WEM, the price of power exchange with the MGs, and the amount of Disco's ramp. The amount of the load interruption, the DGs' power generation, and the results of the DN power flow are considered as the second-stage variables.

The aim of the MG operators (MGOs) is to minimize their operation cost with the scheduling of DERs and the trading power with the Disco. The required input data of the MGOs is the price signal from the Disco and the bids and the technical constraint of DERs. The optimal decisions of the MGOs to trade energy with the Disco are sent to it regarding which the Disco update its mentioned decisions. This process is continued until the equilibrium point among the Disco and the MGs is obtained. In this paper, the equilibrium point is obtained through replacing the proposed non-linear bi-level optimization model into a mixed-integer linear programming (MILP) using the Karush-Kuhn-Tucker (KKT) conditions and dual theory approach. The details of the mathematical modeling are described in the next section.

## III. MATHEMATICAL FORMULATION

The proposed bi-level optimization problem in this paper is described as follows where the operation problem of the Disco and the MGs is modeled as the UL and the LL problems, respectively.

### A. Uncertainty modeling

In this paper, the normal and irradiance probability distribution functions (PDFs) are used to model the uncertainties of the DN's load and the output power of PV systems in the operation problem of the Disco [17] in which both PDFs are discretized into three intervals [18]. Then, using the scenario-tree method [19], 9 scenarios are generated which the probability of each scenario ($\pi_s$) is shown in Table I.

TABLE I
THE NUMBER OF SCENARIO AND THEIR PROBABILITY OF OCCURRENCE

| # Scenario | 1 | 2 | 3 | 4 | 5 | 6 | 7 | 8 | 9 |
|---|---|---|---|---|---|---|---|---|---|
| Probability (%) | 3 | 6 | 12 | 6 | 12 | 42 | 1 | 2 | 7 |

### B. UL problem: Disco

**B.1. Objective function:** The profit of the Disco is modeled as eq. (1). The first term of this equation is used to show the revenue of the Disco from power exchange with the MGs. when $P_{j,i,t}^{MG} > 0$, the Disco sells energy to the MGs and the Disco purchases energy from the MGs for ($P_{j,i,t}^{MG} < 0$). The cost of purchased power from the WEM is modeled in the second term.

The third term shows the amount of penalty implemented from the ISO to eliminate the sharp ramps of the purchased power from the WEM. The fourth term includes the load interruption cost, the cost of purchased power from the DGs, and the revenue from selling energy to DN loads, respectively.

$$Max \left\{ \begin{array}{l} \sum_{t=1}^{T}\sum_{j=1}^{J}\sum_{i=1}^{M}(P_{j,i,t}^{MG})\rho_t^{LEM} - \sum_{t=1}^{T}(P_{i,t}^{E}\big|_{i=1})\rho_t^{WEM} - \sum_{t=1}^{T}\Delta_t^F \rho_t^F \\ -\sum_{s=1}^{S}\pi_s \left( \sum_{i=1}^{N}\sum_{t=1}^{T} P_{i,t,s}^{IL,D}\rho_t^{IL,D} + P_{g,i,t,s}^{DG,D}\rho_g^{DG,D} - P_{i,t,s}^{D}\rho_t^{D} \right) \end{array} \right\} \quad (1)$$

**B.2. DN demand-supply balance:** Eq. (2) describes the power balance of the DN in each bus where the sum of the purchased power from the WEM, the amount of IL and the power generation of DGs and PVs minus the power trading with the MGs and the amount load consumption is equal to the power loss and the power flow of the lines connected to the bus.

$$P_{i,t}^{E}\big|_{i=1} + P_{i,t,s}^{IL,D} + \sum_{g=1}^{G} P_{g,i,t,s}^{DG,D}\big|_{i \in O} + \sum_{k=1}^{K} P_{k,i,t,s}^{pv}\big|_{i \in P}$$
$$-\sum_{j=1}^{J} P_{j,i,t}^{MG}\big|_{i \in M} - P_{i,t,s}^{D} = \sum_{(i,h) \in L} 0.5\left(P_{i,h,t,s}^{flow} + P_{i,h,t,s}^{loss}\right) \quad \forall i,t,s \quad (2)$$

**B.3. DN constraints:** Eqs. (3)-(4) illustrate the power flow, including active power between buses i and h and the amount of respected power losses linearized using the proposed method in [20]. Eqs. (5)- (6) show the current magnitude between buses i and h as well as its upper and lower limitations. Eq. (7) represents the maximum and the minimum permissible voltage magnitude at each DN bus.

$$P_{i,h,t,s}^{fm} + P_{i,h,t,s}^{to} = R_{i,h} I_{i,h,t,s}^2 \quad \forall i,h \in L,t,s \quad (3)$$

$$P_{i,h,t,s}^{fm} - P_{i,h,t,s}^{to} = R_{i,h}(V_{i,t,s}^2 - V_{h,t,s}^2)/Z_{i,h}^2 \quad \forall i,h \in L,t,s \quad (4)$$

$$I_{i,h,t,s} = (V_{i,t,s} - V_{h,t,s})/Z_{i,h} \quad \forall i,h \in L,t,s \quad (5)$$

$$-\overline{I}_{i,h} \leq I_{i,h,t,s} \leq \overline{I}_{i,h} \quad \forall i,h \in L,t,s \quad (6)$$

$$\underline{V}_i \leq V_{i,t,s} \leq \overline{V}_i \quad \forall i \in N,t,s \quad (7)$$

**B.4. DG constraints:** The upper and lower bounds of the DGs output power are presented in Eq. (8). Moreover, the ramp rate characteristics of the DGs are limited by Eqs. (9)-(12).

$$\underline{P}_g^{DG} \leq P_{g,i,t,s}^{DG,D} \leq \overline{P}_g^{DG} \quad \forall g,i \in O,t,s \quad (8)$$

$$P_{g,i,t,s}^{DG,D} - P_{g,i,t-1,s}^{DG,D} \leq RUP_g^{DG} \quad \forall g,i \in O, t>1,s \quad (9)$$

$$P_{g,i,t,s}^{DG,D} - P_g^{DG,ini} \leq RUP_g^{DG} \quad \forall g,i \in O, t=1,s \quad (10)$$

$$P_{g,i,t-1,s}^{DG,D} - P_{g,i,t,s}^{DG,D} \leq RDN_g^{DG} \quad \forall g,i \in O, t>1,s \quad (11)$$

$$P_g^{DG,ini} - P_{g,i,t,s}^{DG,D} \leq RDN_g^{DG} \quad \forall g,i \in O, t=1,s \quad (12)$$

**B.5. IL constraint:** In Eq. (13), the amount of load interruption is limited.

$$0 \leq P_{i,t,s}^{IL,D} \leq \overline{P}^{IL,D} \quad \forall i,t,s \quad (13)$$

**B.6. Disco power and price trading constraint:** Eq. (14) reveals the maximum and minimum limitations of the purchased power from the WEM and Eq. (15) restricts the power exchange price between the Disco and the MGs.

$$0 \leq P_{i,t}^{E} \leq \overline{P}^{E} \quad \forall i=1,t \quad (14)$$

$$0 \leq \rho_t^{LEM} \leq \overline{\rho}^{LEM} \quad \forall t \quad (15)$$

**B.7. Flexibility constraints:** The optimal limitations of the Disco's ramp on the purchased power from the WEM at each time step will be specified in the optimization problem using Eqs. (16)-(17). It means that the Disco decides on the maximum amount of its ramp based on the provided MGs' ramp, as well as the penalty received from the WEM.

$$-\Delta_t^F \leq P_{i,t}^E - P_{i,t-1}^E \leq \Delta_t^F \quad \forall i=1,t \quad (16)$$

$$-\Delta_t^F \leq \sum_j \Delta_{j,i,t}^{MG} \leq \Delta_t^F \quad \forall i \in M,t \quad (17)$$

Thus, the variables set of the UL problem is as follows:
$$V^D = \left[ p_t^{E,in}, p_{i,t,s}^{IL,D}, p_{g,i,t,s}^{DG,D}, \rho_t^{LEM}, \Delta_t^F, p_{i,h}^{fm}, p_{i,h}^{to}, I_{i,h}, V_{i,t} \right].$$

**C. LL problem: MGs**

Eqs. (18)-(32) are related to the operation problem of MGs.

**C.1. Objective function:** The operation cost of the MGs is minimized using Eq. (18) consists of three terms. The first term is the cost/revenue of purchased/sold from/to with the Disco. The next two terms illustrate the operation cost of the DGs and the cost of load interruption, respectively.

$$min \left\{ \sum_{t=1}^{T} \left( \rho_t^{LEM} P_{j,i,t}^{MG} + C_j^{DG} P_{j,t}^{DG} + C_{j,t}^{IL} P_{j,t}^{IL} \right) \right\} \quad (18)$$

**C.2. The demand-supply balance:** The sum of the trading power with the Disco, the power generation of DG, the power generation of PV, and the amount of discharging power of energy storage is equal to the amount of MGLs and the charging power of the energy storage as modeled by Eq. (19).

$$P_{j,i,t}^{MG} + P_{j,t}^{DG} + P_{j,t}^{PV} + P_{j,t}^{IL} + P_{j,t}^{dch} = P_{j,t}^{demand} + P_{j,t}^{ch} \quad \forall j,i \in M,t \quad (19)$$

**C.3. MG power trading constraint:** Eq. (20) shows the allowed range of exchanging power between the Disco and the MGs.

$$-\overline{P}_j^{MG} \leq P_{j,i,t}^{MG} \leq \overline{P}_j^{MG} \quad \forall j,i \in M,t \quad (20)$$

**C.4. DG constraints:** The upper and lower bound as well as the ramp rate characteristics of the DGs are limited by Eqs. (21)-(23).

$$\underline{P}_j^{DG} \leq P_{j,t}^{DG} \leq \overline{P}_j^{DG} \quad \forall j,t \quad (21)$$

$$P_{j,t}^{DG} - P_{j,t-1}^{DG} \leq RUP_j^{DG}, \; P_{j,t-1}^{DG} - P_{j,t}^{DG} \leq RDN_j^{DG} \quad \forall j,t>1 \quad (22)$$

$$P_{j,t}^{DG} - P_j^{DG,ini} \leq RUP_j^{DG}, P_j^{DG,ini} - P_{j,t}^{DG} \leq RDN_j^{DG} \quad \forall j,t=1 \quad (23)$$

**C.5. MGLs interruption constraint:** Eq. (24) describes the permissible amount of load interruption.

$$0 \leq P_{j,t}^{IL} \leq \overline{P}_j^{IL} \quad \forall j,t \quad (24)$$

**C.6. ES constraints:** The amount of charging and discharging power of each ES is limited using (25)-(28). The minimum and maximum limitations of stored energy in the ESs are indicated in Eq. (26). Equations (27)-(28) demonstrate the dynamic behaviour of the ESs at each time step, which depends on the amount of stored energy in the previous time step and charging/discharging power in the present time step [21].

$$0 \leq P_{j,t}^{ch} \leq \overline{P}_j^{ES} \quad , \quad 0 \leq P_{j,t}^{dch} \leq \overline{P}_j^{ES} \quad \forall j,t \quad (25)$$

$$\underline{E}_j \leq E_{j,t} \leq \overline{E}_j \quad \forall j,t \quad (26)$$

$$E_{j,t} = E_{j,t-1} + \eta_j^{ch} P_{j,t}^{ch} - P_{j,t}^{dch}/\eta_j^{dch} \quad \forall j, t > 1 \quad (27)$$

$$E_{j,t} = E_j^{ini} + \eta_j^{ch} P_{j,t}^{ch} - P_{j,t}^{dch}/\eta_j^{dch} \quad \forall j, t = 1 \quad (28)$$

**C.7. MGs' ramp constraint:** Eq. (29) illustrates the amount of MGs' ramp provided using the dynamic behavior of their power exchange with the Disco during the operation time horizon.

$$P_{j,i,t}^{MG} - P_{j,i,t-1}^{MG} = \Delta_{j,i,t}^{MG} \quad \forall j, i \in M, t \quad (29)$$

Thus, the variables set of the LL problems is as follows:

$$V^{MG} = \left[ P_{j,i,t}^{MG}, P_{j,t}^{DG}, P_{j,t}^{IL}, P_{j,t}^{ch}, P_{j,t}^{dch}, E_{j,t}, \Delta_{j,i,t}^{MG} \right]$$

*D. MILP model*

The proposed non-linear bi-level problem is transformed into a non-linear mathematical program with equilibrium constraints (MPEC) model through replacing the LL problem with its KKT conditions which this approach is described in [22, 23]. Then, the non-linear term of the objective function, i.e., $P_{j,i,t}^{MG} \rho_t^{LEM}$, is replaced with the linear terms using the dual theory approach [24] regarding which the MPEC model is transformed into a MILP one.

## IV. NUMERICAL RESULTS

*A. Data*

The effectiveness of the proposed model is investigated on the IEEE 33-bus test system, including three MGs, four DGs, and five PV systems, as shown in Fig. 2. The input data of the DN, DGs, as well as MGs, are given in [22, 25]. The WEM price, the forecasted power generation of PV systems, and the distribution network load (DNL) are extracted from [29]. Also, the DNL interruption contract price is given in Fig. 3. In addition, the penalty and the selling power price to DNL are 140% and 120% of the WEM price, respectively. Also, the contract price for the ILs of the MGs is 80% as high as that for the DNL interruption. The maximum amount of load interruption associated with both the DN and MGs are 30% and 10% of relevant total loads in each time step, respectively. Other necessary data including the maximum limit for the purchased power from the WEM as well as the power exchange between the Disco and the MGs, can be accessible in [22]. The proposed MILP model is solved by CPLEX 12.0 under the GAMS environment, running on a personal computer powered by Core i7 CPU and 8 GB RAM and also, the outcomes of scenario 6 are picked out for analyzing in the next part.

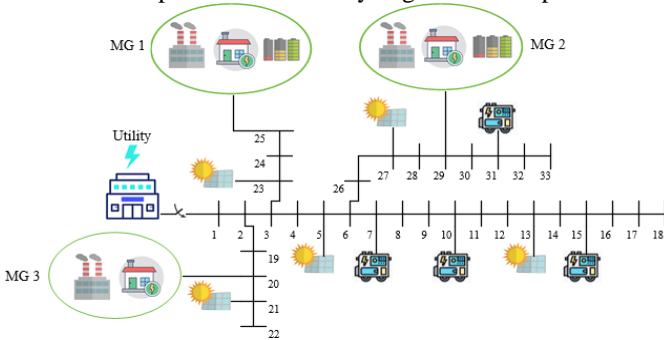

**Fig. 2.** The structure of the modified IEEE 33-bus radial distribution network

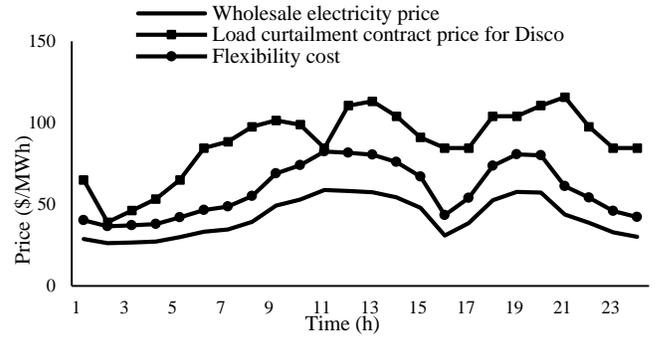

**Fig. 3.** The WEM and DNL interruption prices

*B. Results*

In this sub-section, the operation results of the Disco and the MGs are investigated.

*B.1. Results without flexibility constraints*

The operation results of the Disco and the MGs in the case without the flexibility constraints are presented in Table I and Figs. 4, 5, and 7. As shown in Figs. 4 and 5, the amount of purchased power from the WEM is 194.88MW and maximum amount of ramp up and ramp down in the net load of the Disco, which the Disco purchases it from the WEM, are equal to 8.05 MW/h and -5.80 MW/h in the operation time period. The price of power exchange between the Disco and the MGs is given in Table I. This price depends on the WEM price, the bids of MGs' resources (i.e., DGs and ILs) as well as the dynamic behaviors of the DGs and ESs. In hour 1, the price is cleared at the maximum capacity price, which resulted from the ramp-up limitation of DGs. The price is equal to the bid of MG3's DG in hours 2-7 and 24 and it is equal to the bid of MG2's DG in hours 12-16. Moreover, it is equal to the bids of ILs in the MGs in hours 17-20 and in the other hours, the dynamic behavior of the ESs in the MGs determines the price. The amount of load interruption by the IL and the output power of the DGs in the DN is 8.07MW and 44.99MW, respectively.

*B.2. Results with flexibility constraints*

In this case, the penalty mechanism by the ISO to decrease the ramp of Purchased power of the Disco from WEM is applied to the model regarding which the new price of power exchange among the Disco and the MGs is determined as given in Table 1. Moreover, the operation results of the Disco and the MGs are shown in Figs. 4-6 and 8. Fig. 6 illustrates the optimal amount of ramp of the purchased power from the WEM and the desired amount of ramp for each MG. As mentioned before, the aggregation ramps in MGs should be equal to or lower than the ramp of the purchased power by the Disco from the WEM in each time step. This flexibility-based decision of the Disco and the MGs decreases the ramp of Purchased power from the WEM so that the maximum amount of ramp up and ramp down are 4.55 MW/h and -1.19 MW/h, respectively. Moreover, in this case, the total amount of purchased power from the WEM decreases from 194.88 MW to 183.73 MW in the operation time period. Instead, the amount of load interruption and the output power of the DGs in the DN increases from 8.07 MW and 44.99 MW to 11.81 MW and 49.21 MW, respectively.

Changing the price of power exchange by the Disco leads to rescheduling the MG's resources so that using the DERs in the MGs increases from 370.345 MW to 373.587 MW to provide more flexibility for the system. This decision leads to decreasing the purchased power of the MGs from the Disco from 34.61 MW to 28.49 MW.

The decisions of the Disco and the MGs to meet their demands regarding the price of power exchange among them in the cases without flexibility and with flexibility are shown in Figs. 7 and 8. Considering the flexibility constraints, the Disco and the MGs have three strategies in the operation time period to decrease the ramp rate of the purchased power of the Disco from the WEM as follows:

- Decreasing the purchased power from the WEM:

Fig. 4 indicates that considering flexibility constraints, the Disco purchases lower power from the WEM in hours 1-7, 16-20. This shortage of power is supplied by either increasing the purchased power from the IL and DGs in the DN or changing the amount of power exchange with the MGs. The purchased power from the ILs and DGs connected to the DN increases 2.79 MW and 3.58 MW, respectively in comparison with the case without flexibility. Moreover, the Disco encourages MGs to sell/purchase more/less power to/from the Disco by changing the amount of price offered to the MGs. For instance, the MG1 increases the usage of its resources to decrease/increase the purchased/sold power from/to the Disco in hours 2-6. The MG2 decides to purchase more power from the Disco since the amount of offered price by the Disco in hour 8 is less than the bids of its resources. Also, since the price of power exchange is higher than the bids of MG3's resources in hours 2-7, this MG prefers to increase the power from its sources to sell more power to the Disco.

The purchased power from the WEM by the Disco is also decreased in hours 16-20 through changing the Disco's offers to the MGs. The MG1 prefers to decrease its purchased power from the Disco in hours 16-20 regarding which it discharges the energy storage in these hours. Also, in hours 17-20, with a price of equal to the offers of the Disco, the MG1 increases the amount of ILs. The MG2 changes its role from a consumer to a producer player based on the increase of the Disco's offer in hour 16. Moreover, the amount of sold power to the Disco increases by MG2 in hours 17-20. On the other hand, the MG3 decides to increase the amount of ILs with a price of equal to the offers of the Disco to sell more power to the Disco in hours 18-19.

- Increasing the purchased power from the WEM

Figs. 4 and 5 illustrate that the Disco purchases more power from the WEM to avoid paying the penalty due to its ramp in hours 8-12, 14-15, and 22-24. Thus, the Disco decreases either the purchased power from the DERs in the DN or its offer to the MGs, apart from hour 22, to motivate them to purchase more power from the Disco. As shown in Figs. 7 and 8 (b1, b2), the MG1 decreases the amount of sold power to the Disco in hours 8-9, 11, 23, and 24, and also this MG changes its role from a producer to a consumer in hour 22 to purchase power from the Disco. Figs. 7 and 8 (c1, c2) reveal that the MG2 decides to decrease the sold power to the Disco based on a reduction in the bid of the Disco in hours 10-12 and 14-15. On the other hand, this MG would purchase power as a consumer from the Disco in hours 8-9 and 23.

TABLE I
THE LEM PRICE BEFORE/AFTER CONSIDERING FLEXIBILITY CONSTRAINTS

| Time (h) | LEM price ($/MWh) Before flexibility | LEM price ($/MWh) After flexibility | Time (h) | LEM price ($/MWh) Before flexibility | LEM price ($/MWh) After flexibility |
|---|---|---|---|---|---|
| 1 | 90 | 90 | 13 | 40 | 41.39 |
| 2 | 35 | 38.61 | 14 | 40 | 40 |
| 3 | 35 | 37 | 15 | 40 | 40 |
| 4 | 35 | 35 | 16 | 40 | 55 |
| 5 | 35 | 35 | 17 | 65 | 65 |
| 6 | 35 | 35 | 18 | 80 | 80 |
| 7 | 35 | 40 | 19 | 80 | 80 |
| 8 | 43.10 | 38.61 | 20 | 85 | 85 |
| 9 | 38.9 | 38.61 | 21 | 49 | 55 |
| 10 | 43.68 | 42.78 | 22 | 44.41 | 55 |
| 11 | 44.32 | 40 | 23 | 31.59 | 23.72 |
| 12 | 40 | 38.61 | 24 | 35 | 26.28 |

The MG3 decreases selling power to Disco in hours 23-24 based on the offers of the Disco (see Figs. 7 and 8 (d1, d2)). In hour 22, the Disco makes an affordable decision due to the flexibility constraints so that it increases the offer to the MGs based on the dynamic behavior of the ES in the MG1 and the ramp-rate limitations of the MGs' DGs. Hence, the summation of power trading with the MGs decreases from -3.88 MW to -2.153MW which satisfies the required ramp in the WEM.

- No change in the purchased power from the WEM:

As shown in Figs. 4, 7, and 8 (a1 and a2), the purchased power from the WEM by the Disco maintains unchanged in hours 13, 21; while, the price of power exchange increases. The main reason is that the Disco, based on its behavior in these hours, prefers to change its offers to the MGs to increase its profit. Also, the Disco changes the amount of purchased power from the IL and DGs in the DN to achieve an efficient result.

*C. Discussion*

In the case without flexibility constraints, the Disco's profit is 7849.32$, and the operation cost of MGs 1, 2, and 3 are 4180.97$, 4415.195$, and 4977.885$, respectively. Considering the flexibility constraints in the decision-making problem of the Disco and the MGs changes their operation strategies which lead to changing the Disco's profit and the cost of the MGs. In this case, the Disco's profit is 6025.567$, and the operation cost of MGs 1, 2, and 3 are 4180.83$, 4199.43$, and 5435.59$, respectively. Therefore, the amount of profit reduction of the Disco is 1823.753$ which is considered as the lost revenue of the Disco which can be paid by the ISO. In the market-based mechanism, the ISO pays the cost to the FRPs' providers to meet its required ramp requirements. However, when this mechanism cannot be implemented in some networks, the ISO should pay the cost to the large consumers such as the Discos to mitigate the amount of their ramps in their purchased power from the WEM.

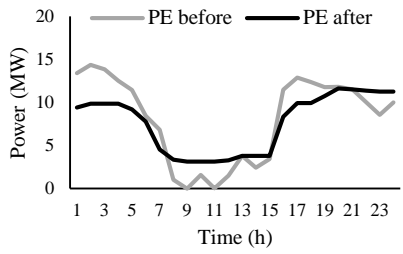

**Fig. 4.** Purchased power from the WEM before/after considering flexibility constraints

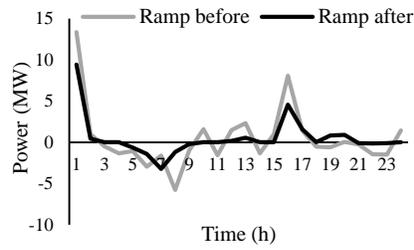

**Fig. 5.** Disco's ramps in purchased power from WEM before/after considering flexibility constraints

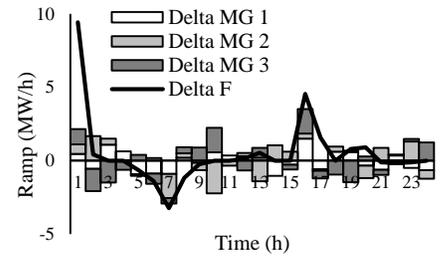

**Fig. 6.** Specified MGs' ramp

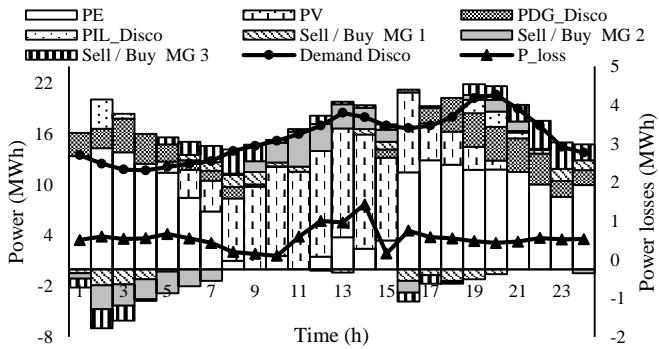

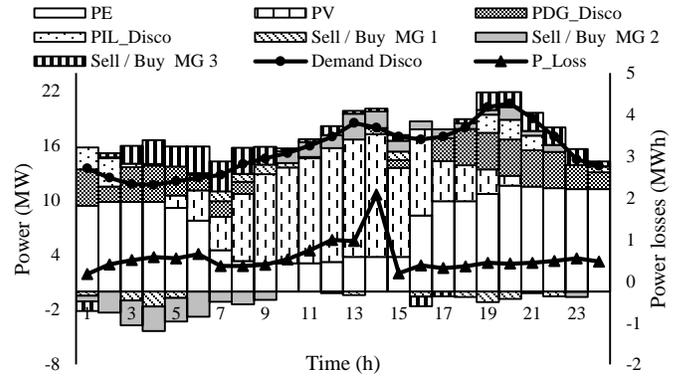

(a2) Disco

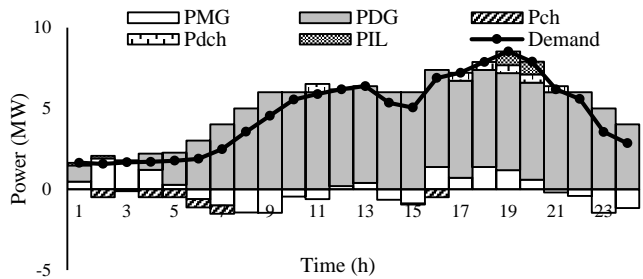

(b1) MG 1

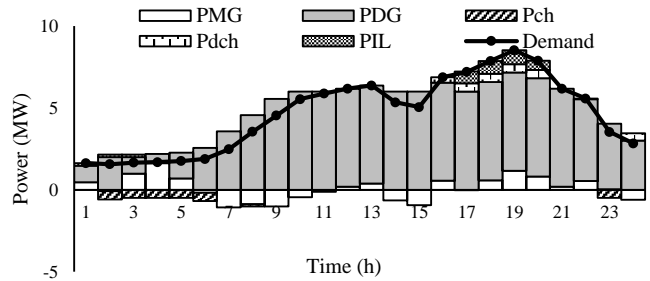

(b2) MG 1

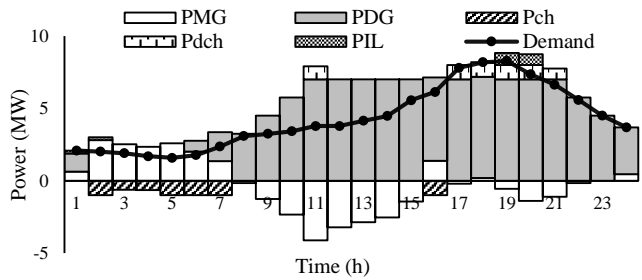

(c1) MG 2

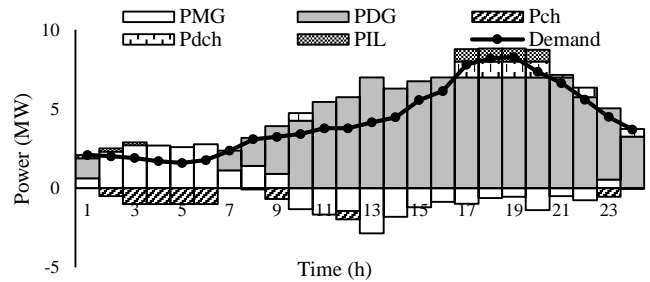

(c2) MG 2

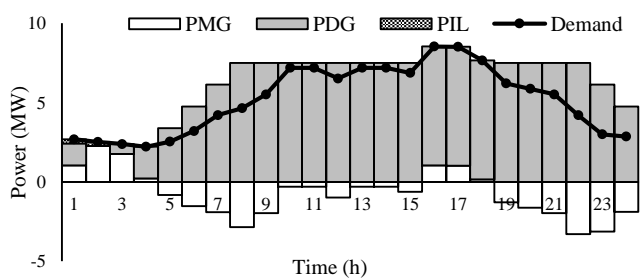

(d1) MG 3

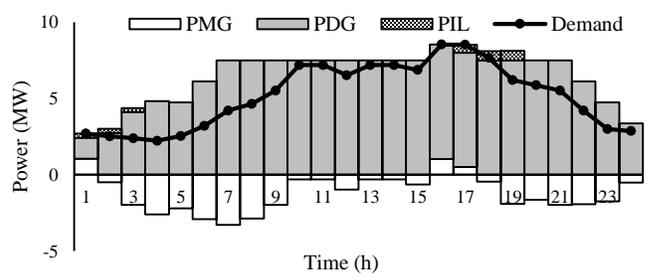

(d2) MG 3

**Fig. 7.** Demand-supply balance in the MGs without flexibility constraints

**Fig. 8.** Demand-supply balance in the MGs with flexibility constraints

## V. Conclusion

The flexibility requirement of the system can be provided through decreasing the ramp rate of the purchased power of the Disco from the WEM regarding two main reasons; 1) Because of the high amount of DN's demand and the power generation of RESs, especially the PV systems in DNs, the purchased power of the Disco from the WEM is an important factor in the 25 net load of the system, and 2) the Disco can use the flexible energy resources in its network to decrease the ramp rate of the purchased power from the WEM. In this paper, a flexibility constrained bi-level two-stage stochastic approach is used to decrease the ramp rate of the purchased power by the Disco from the WEM using the flexible energy sources operated by the Disco and the MGs as the UL and LL decision makers, respectively. The developed model is a non-linear bi-level problem that is transformed into a MILP problem using the KKT conditions and the dual theory. The results of applying the model on a test system with three MGs shows that the maximum amount of ramp-up and ramp-down of Disco's purchased power from WEM decrease from 8.05 MW/h and -5.80 MW/h to 4.55 MW/h and -1.19 MW/h, respectively.